# Sunspot Catalogue of the Observatory of the University of Coimbra (1929–1941)

V.M.S. Carrasco[1,2], J.M. Vaquero[2,3], M.C. Gallego[1,2], A. Lourenço[4], T. Barata[4], J.M. Fernandes[4,5,6]

[1] Departamento de Física, Universidad de Extremadura, 06071 Badajoz, Spain [e-mail: vmscarrasco@unex.es]

[2] Instituto Universitario de Investigación del Agua, Cambio Climático y Sostenibilidad (IACYS), Universidad de Extremadura, 06006 Badajoz, Spain

[3] Departamento de Física, Universidad de Extremadura, 06800 Mérida, Spain

[4] Centro de Investigação da Terra e do Espaço da Universidade de Coimbra, P-3040-004 Coimbra, Portugal

[5] Observatorio Geofísico e Astronómico da Universidade de Coimbra, P-3040-004 Coimbra, Portugal

[6] Departamento de Matematica da Universidade de Coimbra, Largo D. Dinis, P-3001-454 Coimbra, Portugal

**Abstract:** A sunspot catalogue was published by the Coimbra Astronomical Observatory (Portugal), now named Geophysical and Astronomical Observatory of the University of Coimbra, for the period 1929–1941. We digitalized data included in that catalogue and provide a machine-readable version. We show the reconstructions for the (total and hemispheric) sunspot number index and sunspot area according to this catalogue, comparing it with the sunspot number index (version 2) and Balmaceda sunspot area series (Balmaceda *et al*., J. Geophys. Res. 114, A07104, 2009). Moreover, we also compared the Coimbra catalogue with records made at the Royal Greenwich Observatory. The results demonstrate that the historical catalogue compiled by the Coimbra Astronomical Observatory contain reliable sunspot data and therefore can be considered for studies about solar activity.

## 1. Introduction

The sunspot number series is the most used index to characterize the long-term solar activity (Clette *et al*., 2014; Usoskin, 2017). In fact, sunspot counting is considered as the world's longest-running experiments, exceeding the 400 years in duration (Owens, 2013). Sunspot catalogues provide valuable information about several parameters of solar activity (Lefèvre and Clette, 2014). In addition to the number of single sunspots and sunspot groups, parameters as sunspot areas and heliographic coordinates are commonly included in sunspot catalogues. Currently, a limited number of historical sunspot catalogues are available.

Recently, some sunspot catalogues have been digitized. Carrasco *et al*. (2014) and Lefèvre *et al*. (2016) presented a machine-readable version of the sunspot catalogues published by two Spanish astronomical observatories, Valencia (1920-1928) and Madrid (1914-1920), respectively. Mandal *et al*. (2017) have presented the sunspot area series from the digitized and calibrated Kodaikanal white-light sunspot data for the period 1921-2011. Moreover, Willis *et al*. (2013a, 2013b) and Erwin *et al*., (2013) drew attention to the existence of errors contained in the Royal Greenwich Observatory (hereafter RGO) catalogue. Baranyi, Győri, and Ludmány (2016) described the main characteristics of the available data and online tools for the Debrecen Heliophysical Observatory catalogue. Otherwise, recent works (Lefèvre and Clette, 2014; Carrasco *et al*., 2015) have carried out tasks to merge the information included in the sunspot catalogues to obtain larger homogeneous data series. However, this work of recovering historical solar observations should be continued to fill gaps existing in the temporal coverage of the observations and to correct possible shortcomings in the data series (Lefèvre and Clette, 2014).

The aim of this work is to provide a machine-readable version of the sunspot catalogue published by the Astronomical Observatory of the Coimbra University (hereafter COI) for the period 1929-1941. We also present an analysis of the data including a quality-control. In Section 2, we describe the main characteristics of this sunspot catalogue. The sunspot indices obtained from data of the COI catalogue are presented in Section 3. The comparisons with the RGO catalogue and Balmaceda sunspot area series (Balmaceda *et al*., 2009) are shown in Section 4. The main conclusions of this study are exposed in Section 5.

**2. Coimbra Sunspot Catalogue**

The COI is installed at Coimbra (40°12′ N 8°26′ W), a city located in the central region of Portugal. The first Portuguese astrophysics unit was created in this observatory in 1925 due to the effort of the director, Francisco Miranda da Costa Lobo (Bonifácio, 2017; Leonardo, Martins, and Fiolhais, 2011). Systematic (daily) solar sunspot

Sunspot Catalogue of the Coimbra Astronomical Observatory

observations of sunspots, faculae, filaments, and prominences were performed in the COI from 1926 to 1944 on the Ca II line and Hα. A catalogue was published by the institution including these observations except to those ones made in 1926, 1927, and 1928 (da Costa Lobo, 1929), as well as sunspot observations for the period 1942-1944. Furthermore, sunspot and faculae observations made in the COI in 1979 were also published. More recent solar images made from the same instrument (but using a CCD camera) are available at BASS2000 - Solar Survey Archive (http://bass2000.obspm.fr) since 2007. The responsible of solar observations at the COI was the chief observer José António Madeira and his assistant Adelino Pessoa. Figure 1 shows an example page of sunspot observations included in this catalogue. In this sunspot catalogue, the following information can be consulted: i) day and hour of the observation, ii) number of the sunspot group assigned by the COI, iii) number of single sunspots in that group, iv) number of the facula assigned by the COI, v) date when phenomenon was observed for the first time, vi) temporal duration from the first occurrence of the phenomenon, heliographic vii) latitude and viii) longitude limits of the groups, iv) area of the sunspot umbra, and x) total area of the sunspot. Note that areas are given in millionths of solar hemisphere, and the numbers assigned to the sunspot groups are counted from 1929 January 1, beginning with those appearing to the East and North. Moreover, limits of the groups for the heliographic longitude are given from the central meridian, being positive for the East and negative for the West in the case of longitudes, and for the heliographic latitude from the solar equator, being positive for the North and negative for the South.

| 74 | | | ANAIS DO OBSERVATÓRIO ASTRONÓMICO | | | | | | | |
|---|---|---|---|---|---|---|---|---|---|---|
| Janeiro | | | | MANCHAS | | | | | | 1929 |
| Época T. M. G. Loc. obs. | N. | n. | N. F. | D. ap. | D. v. | Lim. Lat. | | Lim. long. | | Sup. S. | Sup. T. |
| 1 9ʰ57ᵐ–10ʰ4ᵐ | 1 | 2 | 2 | 1–I | .. | − 9,5 | − 11,7 | +33,2 | +35,8 | 10 | 64 |
| | 2 | 18 | 4 | » | .. | + 4,8 | + 6,9 | −13,4 | − 23,0 | 97 | 1425 |
| | 3 | 3 | 5 | » | .. | +16,7 | +19,4 | −32,6 | − 36,8 | 8 | 70 |
| | 4 | 1 | 6 | » | 1 d | + 1,8 | | −49,3 | | 5 | 38 |
| 2 10ʰ37ᵐ–44ᵐ | 1 | 4 | 2 | 1–I | .. | − 9,4 | − 23,8 | − 8,4 | +22,1 | 8 | 112 |
| | 2 | 5 | 4 | » | . | + 6,6 | +10,8 | −47,6 | − 51,8 | 51 | 395 |
| | 3 | 2 | 5 | » | .. | +18,2 | +19,2 | −44,4 | − 49,3 | 14 | 137 |
| | 5 | 1 | 1 | 2–I | .. | + 6,4 | | +78,6 | | 24 | 94 |
| 3 10ʰ 9ᵐ–16ᵐ | 1 | 3 | 2 | 1–I | 3 d | −10,0 | − 17,6 | +18,1 | − 21,1 | 7 | 110 |
| | 2 | 3 | 4 | » | .. | + 5,2 | + 8,4 | −40,0 | − 49,2 | 83 | 451 |
| | 3 | 2 | 5 | » | 3 d | +18,4 | +21,0 | −58,4 | − 61,9 | 15 | 165 |
| | 5 | 1 | 1 | 2–I | .. | +12,2 | +15,7 | +73,6 | +75,3 | 19 | 287 |
| 4 9ʰ50ᵐ–57ᵐ | 2 | 9 | 4 | 1–I | .. | + 3,0 | + 7,0 | −42,1 | − 64,2 | 56 | 338 |
| | 5 | 4 | 1 | 2–I | .. | +10,2 | +13,9 | +54,3 | +58,8 | 26 | 208 |
| 5 10ʰ 0ᵐ–7ᵐ | 2 | 2 | 4 | 1–I | .. | + 3,8 | + 8,6 | −55,2 | − 59,9 | 15 | 185 |
| | 5 | 2 | 1 | 2–I | .. | +12,3 | +13,4 | +42,2 | +44,4 | 37 | 264 |
| 6 10ʰ51ᵐ–58ᵐ | 2 | 4 | 4 | 1–I | 6 d | +15,9 | +17,2 | −70,7 | − 75,5 | 19 | 188 |
| | 5 | 3 | 1 | 2–I | .. | +11,1 | +13,5 | +27,7 | +31,3 | 30 | 96 |
| | 6 | 1 | 10 | 6–I | .. | +15,3 | +16,5 | +41,6 | +44,6 | 3 | 19 |
| | 7 | 4 | 16 | » | .. | − 1,9 | − 2,6 | − 9,9 | − 10,4 | 2 | 10 |
| 7 10ʰ28ᵐ–35ᵐ | 5 | 1 | 1 | 2–I | .. | +12,4 | +13,2 | +16,3 | +21,2 | 39 | 218 |
| | 6 | 4 | 10 | 6–I | .. | +18,0 | +26,3 | +29,2 | +29,3 | 4 | 28 |
| | 7 | 4 | 16 | » | .. | − 0,2 | − 3,1 | −19,4 | − 23,9 | 11 | 66 |
| | 8 | 5 | 9 | 7–I | .. | − 9,1 | − 10,6 | +37,3 | +38,9 | 11 | 61 |
| | 9 | 5 | 11 | » | .. | + 8,2 | + 9,7 | −33,2 | − 38,2 | 13 | 92 |
| 8 9ʰ52ᵐ–59ᵐ | 5 | 1 | 1 | 2–I | .. | + 8,8 | +12,1 | + 2,6 | + 5,0 | 20 | 134 |
| | 6 | 2 | 1 | 6–I | 3 | + 9,0 | +11,2 | +15,6 | +18,5 | 5 | 58 |
| | 7 | 2 | 16 | » | 3 | − 2,4 | − 4,2 | −33,5 | − 39,6 | 11 | 104 |
| | 8 | 3 | 9 | 7–I | .. | − 8,9 | − 11,6 | +24,8 | +26,9 | 14 | 129 |
| | 9 | 4 | 11 | » | 2 | + 6,0 | + 7,9 | −47,6 | − 53,0 | 19 | 111 |
| | 10 | 3 | 1 | 8–I | 1 | +11,5 | +12,8 | −12,8 | − 17,2 | 5 | 84 |
| 10 9ʰ57ᵐ–10ʰ4ᵐ | 5 | 1 | 1 | 2–I | .. | +11,1 | +13,0 | −21,9 | − 24,1 | 7 | 83 |
| | 8 | 4 | 9 | 7–I | .. | −12,0 | − 14,1 | − 1,5 | + 1,5 | 10 | 92 |
| | 11 | 4 | 17 | 10–I | .. | +19,1 | +21,7 | +54,7 | +57,0 | 13 | 187 |
| 11 9ʰ38ᵐ–45ᵐ | 5 | 1 | 1 | 2–I | 10 | + 9,9 | +12,6 | −33,8 | − 35,7 | 8 | 55 |

Figure 1. An example page of the Coimbra sunspot catalogue [Source: da Costa Lobo, 1929, page 74].

We have made a machine-readable version of this catalogue and it is publicly available at the website of the Historical Archive of Sunspot Observations (http://haso.unex.es/). Table 1 shows some lines of our machine-readable version which it has the following format:

- First (YEAR), second (MONTH), and third (DAY) columns list the year, month, and day of the observations.

- Fourth (N) and fifth (n) columns indicate numbers assigned by COI to sunspot groups and the number of single sunspots observed in that group.

Sunspot Catalogue of the Coimbra Astronomical Observatory

- sixth and seven columns (LATITUDE LIMITS) give limits of the sunspot groups according to the heliographic latitude, and the eighth and ninth columns (LONGITUDE LIMIT), limits for the heliographic longitude from the central meridian.
- Tenth (UMBRA AREA) and eleventh (TOTAL AREA) columns indicate measurements for the umbra and total area of the sunspot groups.

Table 1. Some example lines of our machine-readable version for the COI catalogue.

| YEAR | MONTH | DAY | N | n | LATITUDE LIMITS | | LONGITUDE LIMITS | | UMBRA AREA | TOTAL AREA |
|---|---|---|---|---|---|---|---|---|---|---|
| 1929 | 1 | 1 | 1 | 2 | -9.5 | -11.5 | 33.2 | 35.8 | 10 | 64 |
| 1934 | 1 | 18 | 432 | 1 | 3.4 | 4.3 | -79.5 | -80.4 | 0 | 32 |
| 1937 | 10 | 6 | 1207 | 11 | 4.4 | 13.2 | -19.5 | -37.7 | 256 | 4886 |

The instrument chosen to perform solar observations in COI was a spectroheliograph similar to the one installed at Meudon Observatory in Paris. In fact, Henri Deslandres, director of the Meudon Observatory in that time, played a key role in the implementation of the equipment at the COI since it was constructed and installed following specifications given by Deslandres (Bonifácio 2017; Leonardo, Martins, and Fiolhais, 2011). The spectroheliography had a coelostat composed by two mirrors with a 0.4-meter diameter each one of them. The coelostat was in an external pavilion and sent the light of the Sun through a small window towards an objective with 0.25 m in aperture and a focal length of 4 m. Then, the solar light beam was deflected before reaching the first slit towards a vertical screen, where an image of the Sun with a 0.4 m was projected. Furthermore, one photographic image of 10 cm in diameter was also obtained. On the other hand, a special device allowed to obtain pictures of the solar region of interest on the 0.4 m projected image to study details of single sunspots. All the details included in these images were transported to a planned image where the solar visible hemisphere was divided into 9 parallels of 10º each one and 36 equal-angles sectors disposed in a radial belt whose peaks are in the centre of the solar disc where the distortion of the image is zero (Mouradian and Garcia, 2007). Then, the determination of heliographic coordinates and sunspot areas were made on a planned image except to those sunspots that, due to its number, allowed to perform the measurements directly in the projected image without loose of quality (Costa Lobo, 1929, pp. 10-16).

In order to identify mistakes in the COI catalogue, we performed a check of the different parameters included in the catalogue. Year, month, and day should be between the logical parameters 1929-1941, 1-12, 1-31, respectively, as well as heliographic latitudes and longitudes within the interval from -90º to +90º. A sunspot should not increase its heliographic longitude in 20º per day and its latitude in 5º. We have also checked possible problems when the extent of sunspot groups is greater than 25º in longitude or 10º in latitude. As in the COI catalogue, a new group number is assigned to the same sunspot group when it again appears on the solar disc in a new rotation, then the numbers assigned to same groups should not be temporarily separated for more than a month. Thus, several mistakes have been found in the COI catalogue. However, some of these errors could not be corrected and still remain in the catalogue. These errors are: i) 211 omissions of parameters (76 cases for latitudes, 77 for longitudes, 54 for umbrae, 2 for total area, and groups were not registered in 2 cases for a given day when they were registered previous and/or subsequent days), ii) 11 errors according to values of heliographic latitudes, iii) 12 cases in values of heliographic longitudes, and iv) 4 mistakes in values of sunspot umbra measurements.

The total number of sunspot observations made at the COI for the period 1929-1941 was 3097. On average, it implies 238 observations per year and a temporal coverage of 65.2% for the entire observation period. Note that no observations were made in January 1931 because the building was closed for works. Figure 2 shows the distribution of the number of individual sunspots within the groups (left panel) and the measured areas (right panel) registered for groups in the COI catalogue. Note that Figure 2 (left panel) only shows the distribution for sunspot groups composed by a number equal or lower than 20 individual sunspots, although groups up to 41 sunspots can be found in the COI catalogue (the number of groups with a number of individual sunspots between 21 and 41 is lower than 15). In the case of area measurements (Figure 2, right panel) only groups with areas lower than 800 millionths of solar hemisphere are represented, although the maximum area for a sunspot group recorded in the COI catalogue, on 6 October 1937, is equal to 4886 millionths of solar hemisphere (the frequency for groups with an area between 800 and 4900 millionths of solar hemisphere is equal to or lower than 25). The smallest sunspot groups according to the single sunspot number and area measurement are the most numerous in the COI catalogue.

Sunspot Catalogue of the Coimbra Astronomical Observatory

Figure 2 also shows that frequency decreases when groups grow both in sunspot number and area occupied on the photosphere.

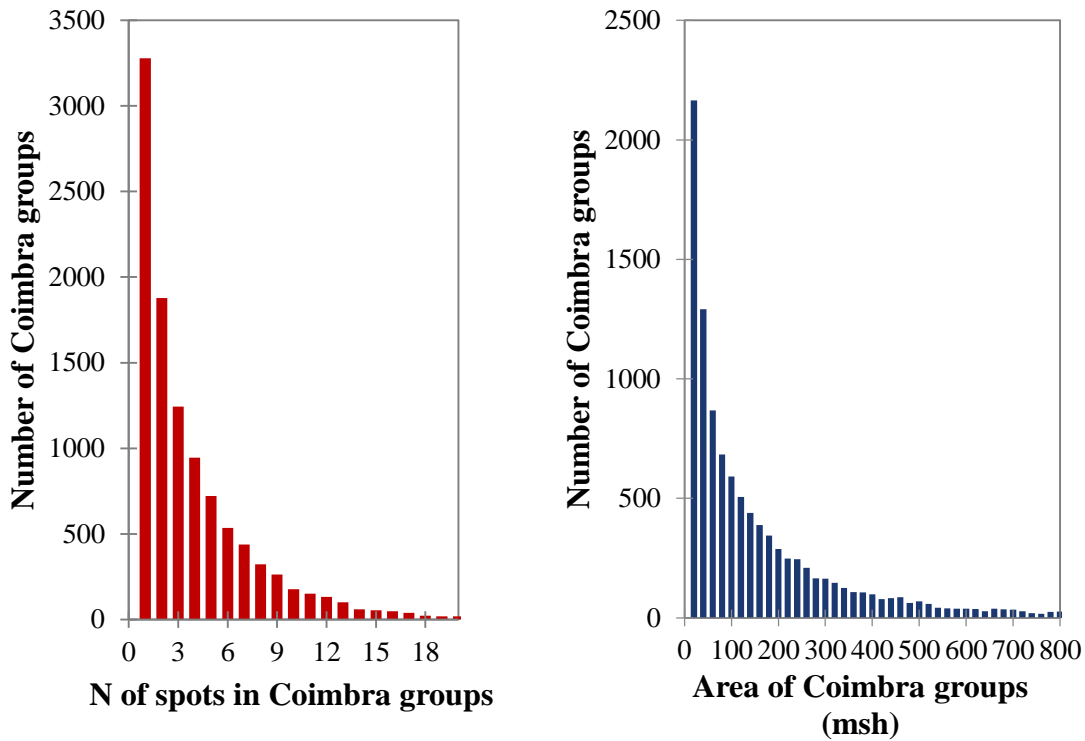

Figure 2. Distribution of (*left panel*) the number of individual spots and (*right panel*) measured areas for groups recorded in the COI catalogue. Area data are represented in bins of 20 millionths of solar hemisphere (msh).

**3. Solar Indices**

**3.1. Sunspot Number**

We compute the sunspot number according to the COI data applying the definition: $CSN = 10G + S$, being $G$ the number of sunspot groups and $S$ the number of single sunspots registered. Figure 3 depicts the temporal evolution of the $CSN$ and the Sunspot Number (version 2, $S_N$) for the period 1929-1941. Data for the $S_N$ were extracted from the Sunspot Index and Long-term Solar Observations website (http://sidc.oma.be/silso/). The observation period approximately spans one solar cycle, *i.e.* from the declining phase of Solar Cycle 16 to that of Solar Cycle 17. Both series set the solar minimum (1933) and maximum (1937) for the Solar Cycle 17 in the same years. Moreover, values of the two series are closer in the solar minimum while those are clearly far in the maximum where the values according to $S_N$ are approximately twice higher than the $CSN$ values.

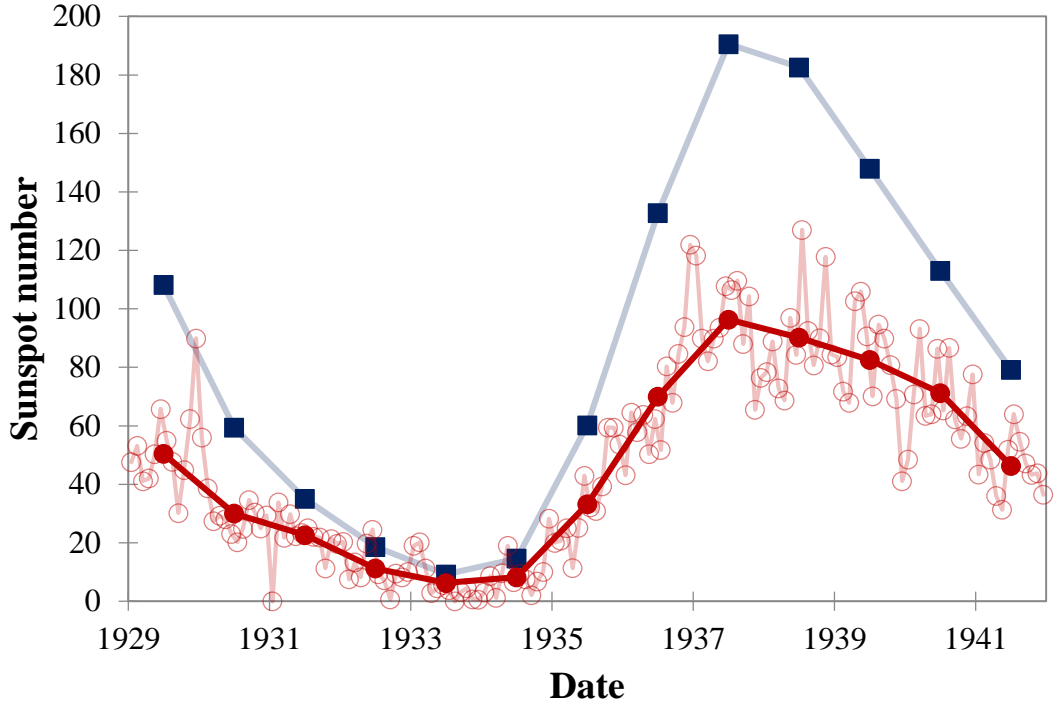

Figure 3. Temporal evolution of the annual Sunspot Number index version 2 (*blue squares*), and monthly (*open red circles*) and annual (*solid red circles*) Coimbra Sunspot Number.

We calculated the calibration factor for the COI by comparing monthly *CSN* and $S_N$ data (Figure 4). The expression for the best linear fit is: $S_N = (1.96 \pm 0.34)\ CSN + (-4.56 \pm 2.00)$, $r = 0.977$, *p*-value < 0.001. If we set the y-intercept of the regression line to zero, we obtained the calibration factor for COI: $k_{CSN} = 1.89 \pm 0.02$. Furthermore, in order to analyse the variability of *CSN* respect to $S_N$, we calculated the ratio between the *CSN* and $S_N$ values (Figure 5). A "zero" values for *CSN* (August 1933) and their corresponding $S_N$ values for those dates have been removed for this calculation. Figure 5 does not show, in general, significant changes in the evolution of the ratio. However, we highlight that the difference between the ratio value for 1929 (0.47) and 1940 (0.63), years with a similar $S_N$ (see Figure 3), is equal to 28.6 % compared to the average ratio (0.56) for the whole period. Since the first year of the publication of the COI catalogue was 1929, this fact could be explained by an improvement over the time in the observation capability of the observers and the observation technique. Moreover, note that the ratios show a greater dispersion in the data around the solar minimum.

Sunspot Catalogue of the Coimbra Astronomical Observatory

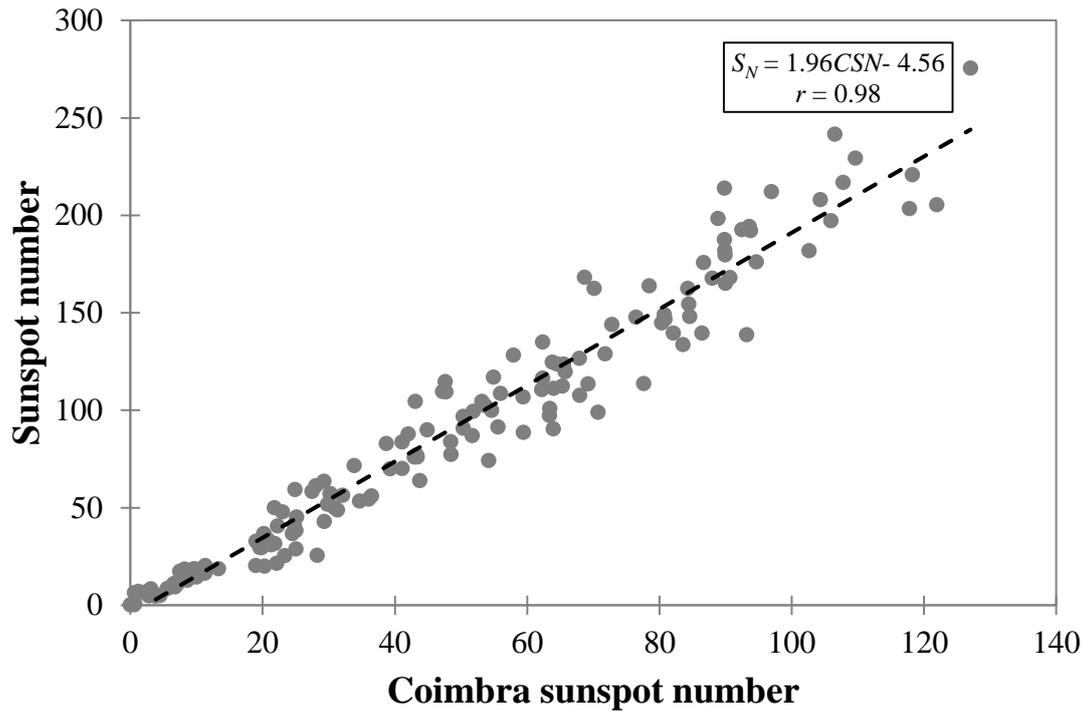

Figure 4. Linear relationship between monthly Coimbra Sunspot Number and Sunspot Number version 2. The best linear-fit equation and correlation coefficient ($r = 0.98$) are shown.

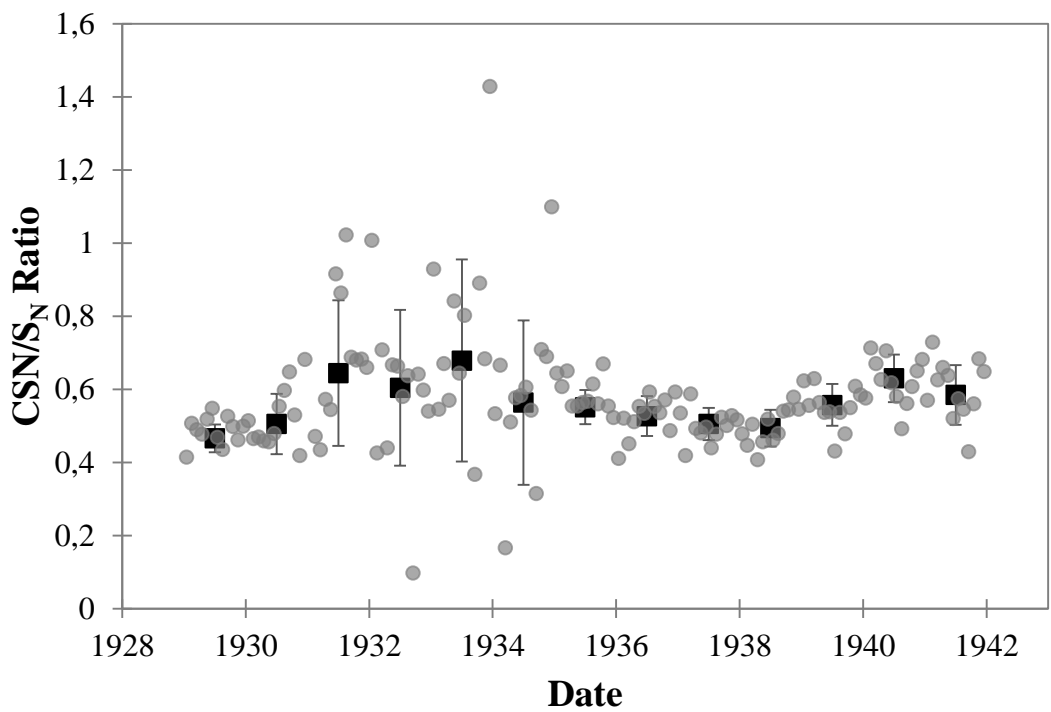

Figure 5. Time evolution of annual (*black squares*) and monthly (*grey circles*) ratios between the Coimbra Sunspot Number and Sunspot Number version 2. The error bars represent one sigma standard deviation.

**3.2. Hemispheric Indices**

In the COI catalogue, the limits of the heliographic latitudes were registered for each sunspot group. Sometimes, one of the limits of the group lies in the Northern Hemisphere (positive latitude) of the Sun and the other one in the Southern (negative latitude). For the purpose of performing an analysis about solar hemispheres, the observations that fulfill this case are considered in the calculations for both hemispheres. Then, the definition of sunspot number index explained in Section 3.1, is independently applied to both hemispheres. Figure 6 represents monthly and yearly values obtained for the *CSN* and the two hemispheres. The solar hemispheric maxima and minima for the Solar Cycle 17 do not match according to the Northern and Southern sunspot number. Considering the Northern Hemisphere, the solar maximum for the Solar Cycle 17 occurs in 1937, as in the *CSN*, while for the Southern Hemisphere it is in 1939. Instead, the solar minimum according to the Southern Hemisphere (1933) matches in time with the *CSN* series but it occurs one year later (1934) in the Northern Hemisphere. Moreover, the Southern Hemisphere seems to be in advanced phase with respect to the Northern Hemisphere. Note that Zolotova *et al*. (2009, 2010) showed from RGO data that the Southern Hemisphere was in advanced phase with respect to the Northern Hemisphere from 1928, when a phase change happened.

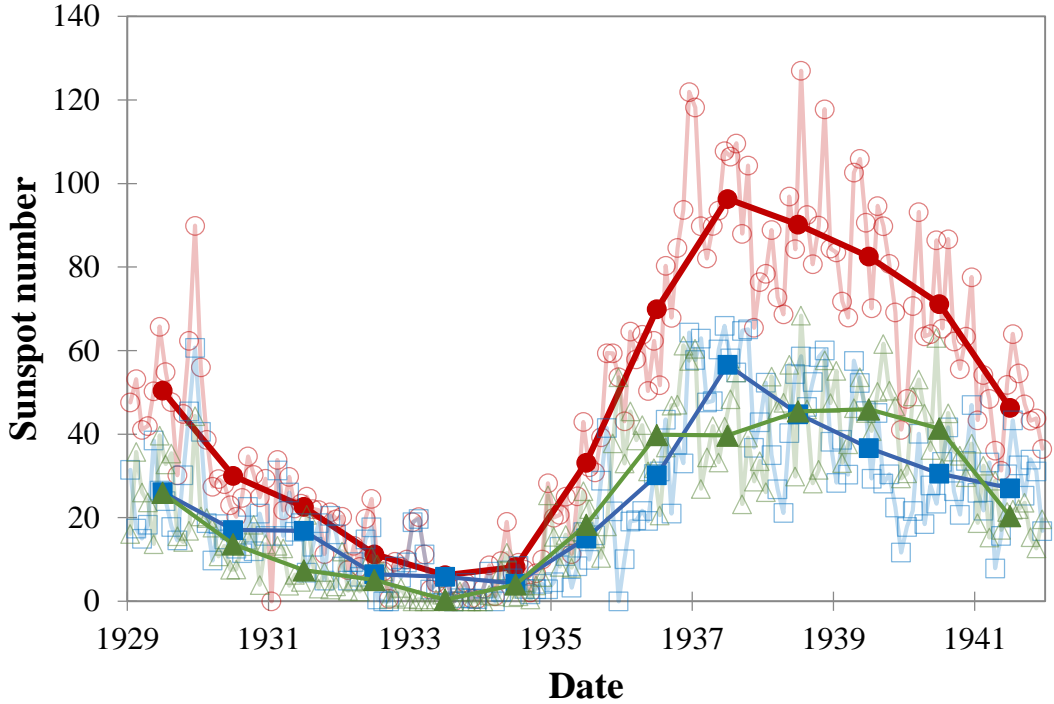

Figure 6. Temporal evolution of the Coimbra Sunspot Number index (*red circles*), the Northern Coimbra Sunspot Number (*blue squares*), and Southern Coimbra Sunspot Number (*green triangles*). *Solid symbols* represent yearly values and *open symbols* represent monthly values.

We have studied the normalized asymmetry through the definition:

$$NA = \frac{G_N - G_S}{G_N + G_S}$$

where *NA* is the normalized asymmetry and $G_N$ and $G_S$ are the number of sunspot groups registered for a given day in the Northern and Southern Hemisphere, respectively. Annual and monthly values were calculated from the average of the daily normalized asymmetry. Figure 7 shows the evolution in time of the normalized asymmetry obtained from the COI catalogue. The average for the normalized asymmetry for the whole period 1929–1941 is positive (0.09), *i.e.* the hemisphere with a greater weight for this period was the Northern Hemisphere. However, the behaviour of this parameter in the Solar Cycle 16 and 17 is different. For the decline phase of the Solar Cycle 16 (1929-1933), along with the first year of Solar Cycle 17 (1934), the Northern Hemisphere clearly dominated in relation to the Southern Hemisphere (*NA* =

0.25). However, the Southern Hemisphere was the dominant hemisphere (*NA* = -0.04) for the remaining period (1935-1941), belonging to the Solar Cycle 17.

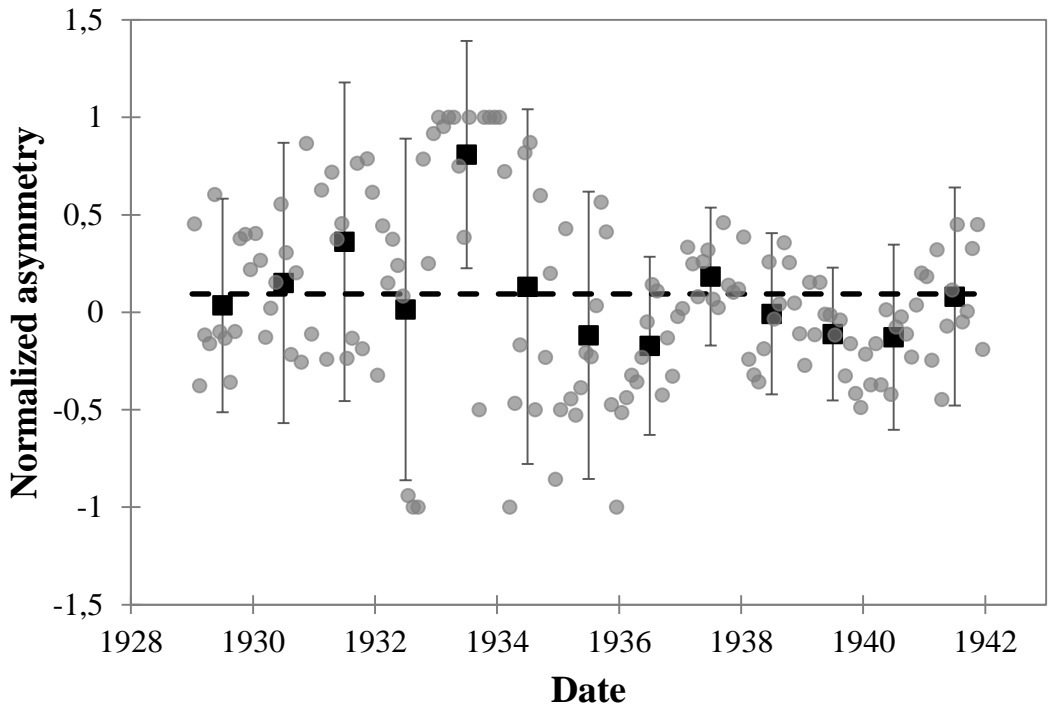

Figure 7. Temporal evolution of the normalized asymmetry according to data recorded in the Coimbra catalogue. *Black squares* and *grey circles* represent annual and monthly values, respectively. The *dashed line* represents the mean annual value of the normalized asymmetry for the entire period 1929–1941. The error bars represent one standard deviation.

### 3.3. Butterfly diagram

The butterfly diagram is a graphical representation of sunspot latitudes observed on the solar disc. In Figure 8, we present the butterfly diagram according to the COI catalogue. Note that, in the COI catalogue, two values in latitude are given for each sunspot group. These values represent the limits of the groups. In order to represent the butterfly diagram, one value in latitude was assigned for each group. We calculated that value as the average from those two limit values recorded in the COI catalogue.

In the butterfly diagram, it can be seen how sunspots appear in lower latitudes as the solar cycle evolves. Values with the higher average latitudes according to the COI catalogue are +38.05º in the North and -36.95º in the South. At the end of the Solar Cycle 16, there are a higher number of sunspot groups and with greater size in the Northern Hemisphere. However, at the beginning of the Solar Cycle 17, the higher number of groups and the larger sunspot groups appeared in the Southern Hemisphere. Then, both the number and size of the sunspot groups are more balanced during the Solar Cycle 17.

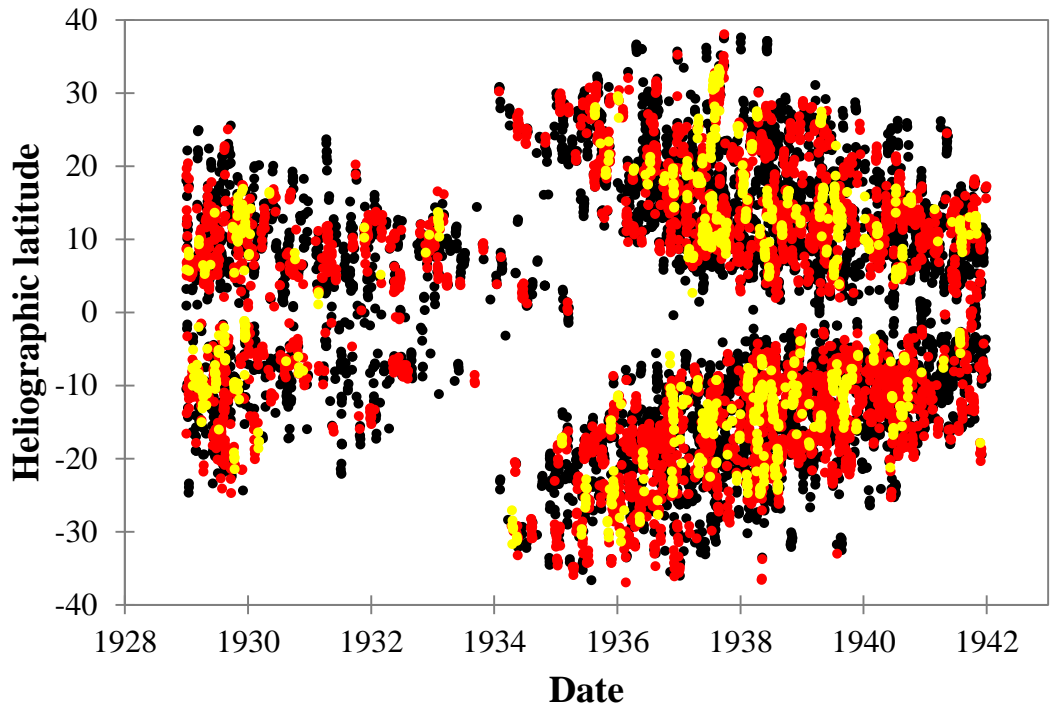

Figure 8. Butterfly diagram according to heliographic latitudes recorded in the Coimbra catalogue. *Colour dots* represent sunspot areas: i) (*black*) smaller than 100 millionths of the solar hemisphere, ii) (*red*) between 100 and 499 millionths of the solar hemisphere, and iii) (*yellow*) larger than or equal to 500 millionths of the solar hemisphere.

## 4. Comparison with other Sources

### 4.1. Comparison with the RGO catalogue

The RGO, with the help of other collaborating observatories, carried out a valuable programme of solar observations during the period 1874-1976 (Willis *et al*., 2013a, 2013b; Erwin *et al*., 2013). Later, the responsibility for the solar observation programme was transferred to Debrecen Heliophysical Observatory. The RGO catalogue is a reference dataset for many studies. For example, Maunder (1904) showed, from the famous butterfly diagram, the migration of heliographic latitudes of sunspots towards the solar equator during the course of the solar cycle using heliographic coordinates from the RGO catalogue for the period 1874-1902. Hoyt and Eddy (1982) used RGO data to develop a model about the total solar irradiance and Balmaceda *et al*. (2009) constructed a homogeneous sunspot area series with an extension about 130 years having the RGO catalogue as a basis.

We compared sunspot groups registered by COI with those included in the RGO catalogue. RGO data used in this work were obtained from the website (https://solarscience.msfc.nasa.gov/greenwch.shtml). To identify the same sunspot group in both observatories, the sunspot group must fulfill some specific constraints in latitude and longitude in both observatories. That is, we consider that a group observed in the COI is the same group observed in the RGO if the latitude and longitude value registered in the RGO for that group lies within the range defined by latitude and longitude value of the COI group plus/minus a certain interval. Longitudes and latitudes of the sunspot groups are defined from the central meridian and solar equator, respectively. We decided that interval employed to identify groups between COI and RGO is ±15º for the longitude and ±5º for the latitude, similar to Carrasco *et al*. (2014).

Applying those criteria, we found that, for the period 1929–1941, 2337 sunspot groups registered in the COI catalogue have corresponding entry in the RGO catalogue, *i.e*. a 96.7% of the total groups registered in the COI catalogue (Table 2). In the case of the RGO, 2884 groups (53.0% of the groups registered in the RGO) have an association with groups registered in the COI catalogue. Considering observation days when sunspot observations were made at both observatories, 2411 sunspot groups (99.8% of the total groups observed by COI) were registered in the COI and 3806 (86.4% of the total groups observed in RGO) were registered in RGO. In 814 cases, one COI group has correspondence with two or more sunspot groups in RGO data, and 803 RGO groups are associated with two or more groups in the COI catalogue. Moreover, the number of observation days in the COI and RGO for the period 1929–1941 was equal to 3097 and 4202, respectively. These numbers indicate a temporal coverage equal to 65.2% and 88.5% for all possible observation days for 1929–1941. The number of the observation days when observations were performed at both observatories is 2715, a 57.2% with respect to all the days of the period 1929–1941.

Table 2. Comparison between sunspot observations made at COI and RGO.

|  | COIMBRA | RGO |
|---|---|---|
| Observation days [%] | 3097 [65.2%] | 4202 [88.5%] |

Sunspot Catalogue of the Coimbra Astronomical Observatory

| Coincident observation days [%] | 2715 [57.2%] | |
|---|---|---|
| No observations at all [%] | 167 [3.5%] | |
| Total number of groups | 2416 | 4407 |
| Groups with correspondence in other [%] | 2337 [96.7%] | 2884 [53.0%] |
| Groups when observations at both [%] | 2411 [99.8%] | 3806 [86.4%] |

We also compared the group number series computed from the COI and RGO data (Figure 9). The yearly values were calculated from the daily average. Figure 9 shows that the behaviour of the solar cycle in both indices is similar but the values of the RGO group number are always greater than the COI group number. Both series set the solar maximum for Solar Cycle 17 in 1937, as the *CSN* and $S_N$. However, there are discrepancies in the occurrence of the solar minimum of Solar Cycle 17. According to the RGO series, the solar minimum occurs in 1934, while it is in 1933 for the COI group number series, as the *CSN* and $S_N$ indices. Furthermore, we can obtain the calibration factor for the COI relatively to the RGO by comparing the monthly average of the group numbers. Thus, the best linear fit is: $G_{RGO} = (1.34 \pm 0.03) G_{COI} + (0.50 \pm 0.14)$, $r = 0.952$, $p$-value $< 0.001$. If we set the y-intercept of the regression line to zero, we obtained the calibration factor for the COI: $k_{CSN} = 1.44 \pm 0.02$.

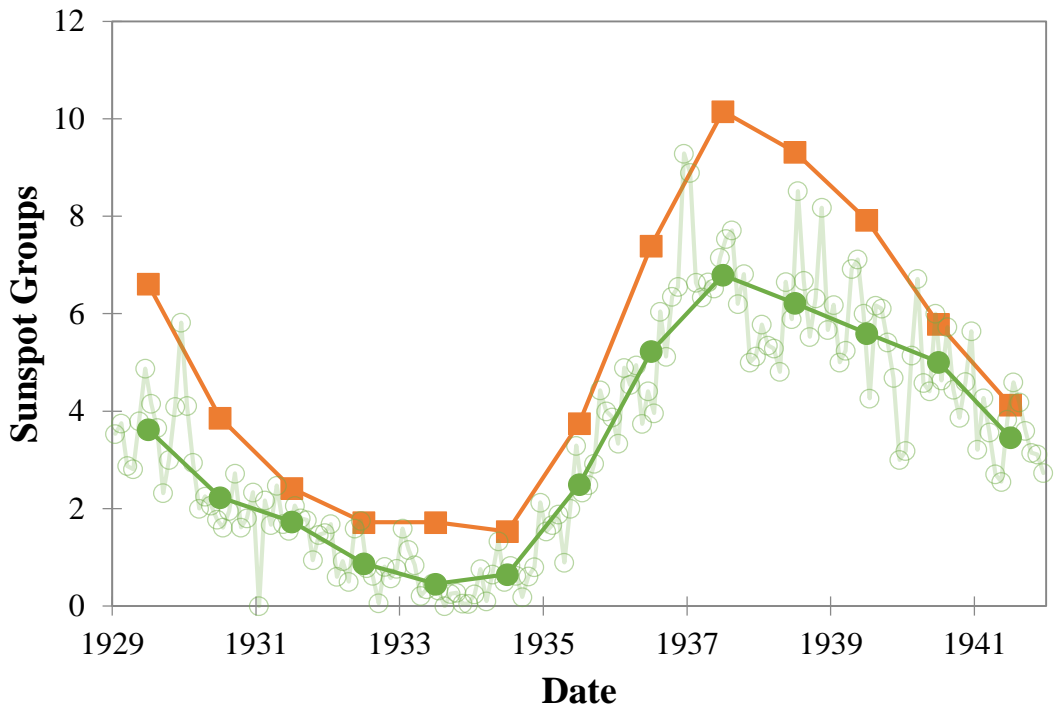

Figure 9. Temporal evolution of the RGO group number (*orange squares*) and monthly (*open green circles*) and annual (*solid green circles*) COI group number.

**4.2. Comparison with Balmaceda Area Database**

The sunspot area series is a valuable index to study the long-term solar activity (Balamaceda *et al*., 2009). We reconstructed the sunspot area series from the COI catalogue and we compared it with the Balmaceda area dataset (Figure 10). In both series, the solar minimum and maximum for Solar Cycle 17 occur in 1933 and 1937, as in *CSN* and $S_N$. The behaviour of both series is similar, but the Balmaceda series presents slightly higher values than the COI series for all the annual values for the period 1929–1941. In addition, we obtained the calibration factor for the COI (*CSA*) in relation to the Balmaceda sunspot area (*BSA*). The equation for the best linear fit comparing the monthly values of both series is: $BSA = (1.19 \pm 0.02)\, CSA + (96 \pm 21)$, $r = 0.974$, *p*-value $< 0.001$. If we set the y-intercept of the regression line to zero, we obtained the calibration factor: $k_{CSN} = 1.27 \pm 0.02$.

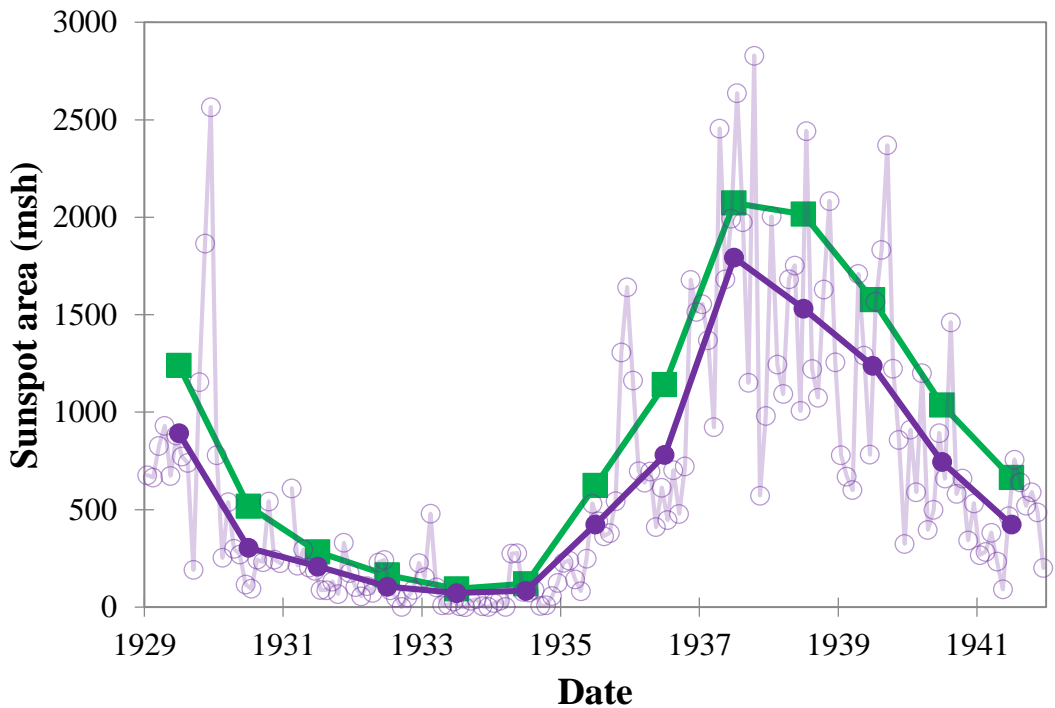

Figure 10. Temporal evolution of the Balmaceda sunspot area (*green squares*) series, and monthly (*open purple circles*) and annual (*solid purple circles*) sunspot area from the Coimbra catalogue. The units of the area measurements are given by millionths of the solar hemisphere.

## 5. Conclusions

We recovered and digitalized data included in the sunspot catalogue published by the Geophysical and Astronomical Observatory of the University of Coimbra (COI) for the period 1929–1941. These data start in the declining phase of the Solar Cycle 16 and finish in the declining phase of the Solar Cycle 17. In this work, we provide a machine-readable version of this catalogue, publicly available at http://haso.unex.es/.

We reconstructed the sunspot number using the Coimbra catalogue (*CSN*) and we obtained a calibration factor with respect to the Sunspot Number (version 2, $S_N$) equal to $k_{CSN} = 1.89 \pm 0.02$. The ratio between the *CSN* and $S_N$ values, in general, does not show significant changes although there is a significant difference in the ratios for 1929 (0.47) and 1940 (0.63), both years with the same solar activity level according to $S_N$. This difference could be explained by an improvement in the quality of the observations in the COI. We also computed the hemispheric sunspot number series. The hemisphere with greater weight for the period 1929-1941 according to the Coimbra catalogue was the Northern Hemisphere, although the Southern Hemisphere slightly dominated in Solar Cycle 17. Moreover, the Southern Hemisphere seems to be in advanced phase with respect to the Northern Hemisphere. We also show the butterfly diagram plotting heliographic latitudes of sunspots recorded in the Coimbra catalogue.

The Coimbra catalogue was compared with other data sources. For the RGO catalogue, the average group number was computed and compared with the COI dataset. We found a similar behaviour and obtained a calibration factor for Coimbra group number equal to $k_{CSN} = 1.44 \pm 0.02$ with respect to the RGO group number. We found a discrepancy between the group number series from both observatories: RGO set the solar minimum for the Solar Cycle 17 in 1934 and the COI in 1933, as *CSN* and $S_N$. Furthermore, we carried out an analysis between sunspot groups registered in the COI and RGO in order to identify the same groups in two observatories. Thus, we found that 96.7% of the total groups registered in Coimbra and 53.0% in the RGO have correspondence in the record by the other observatory. The Coimbra sunspot area series presents a similar behaviour to Balmaceda sunspot area series. The calibration factor for the COI with respect to Balmaceda series is $k_{CSN} = 1.27 \pm 0.02$.

This work demonstrates that sunspot data included in the Coimbra catalogue are reliable and therefore can be employed in solar activity studies. As future work, we will compare areas obtained from an automatic tool developed in the COI to identify sunspots in the drawings with those included in the catalogues. Moreover, other solar activity indices as facular regions, prominences, and filaments can be calculated from the COI data. The recovery of historical sunspot data should be continued in order to obtain more reliable and complete series. In this way, we will understand better the behaviour of solar activity.

**Appendix:**

Bibliographic references containing the original data of the Coimbra Astronomical Observatory are listed in this appendix:

da Costa Lobo, F.M.: 1929, Anais do Observatório Astronómico da Universidade de Coimbra Tomo I, Universidade de Coimbra, Coimbra.

da Costa Lobo, F.M.: 1933, Anais do Observatório Astronómico da Universidade de Coimbra Tomo II, Universidade de Coimbra, Coimbra.

dos Reis, M.: 1936, Anais do Observatório Astronómico da Universidade de Coimbra Tomo III 1931, Universidade de Coimbra, Coimbra.

dos Reis, M.: 1937, Anais do Observatório Astronómico da Universidade de Coimbra Tomo IV 1932, Universidade de Coimbra, Coimbra.

dos Reis, M.: 1940, Anais do Observatório Astronómico da Universidade de Coimbra Tomo V 1933, Universidade de Coimbra, Coimbra.

dos Reis, M.: 1941, Anais do Observatório Astronómico da Universidade de Coimbra Tomo VI 1934, Universidade de Coimbra, Coimbra.

dos Reis, M.: 1942, Anais do Observatório Astronómico da Universidade de Coimbra Tomo VII 1935, Universidade de Coimbra, Coimbra.

dos Reis, M.: 1943, Anais do Observatório Astronómico da Universidade de Coimbra Tomo VIII 1936, Universidade de Coimbra, Coimbra.

dos Reis, M.: 1945, Anais do Observatório Astronómico da Universidade de Coimbra Tomo IX 1937, Universidade de Coimbra, Coimbra.

dos Reis, M.: 1947, Anais do Observatório Astronómico da Universidade de Coimbra Tomo X 1938, Universidade de Coimbra, Coimbra.

dos Reis, M.: 1949, Anais do Observatório Astronómico da Universidade de Coimbra Tomo XI 1939, Universidade de Coimbra, Coimbra.

dos Reis, M.: 1957, Anais do Observatório Astronómico da Universidade de Coimbra Tomo XII 1940, Universidade de Coimbra, Coimbra.

dos Reis, M.: 1965, Anais do Observatório Astronómico da Universidade de Coimbra Tomo XI 1941, Universidade de Coimbra, Coimbra.

**Acknowledgements**

This work was partly funded by FEDER-Junta de Extremadura (Research Group Grant GR15137 and project IB16127) and from the Ministerio de Economía y Competitividad of the Spanish Government (AYA2014-57556-P and CGL2017-87917-P). JF, AB, and AL acknowledge support from the Fundação para a Ciência e a Tecnologia and Centro de investigação da Terra e do Espaço da Universidade de Coimbra Funds (UID/Multi/00611/2013) and ReNATURE (CENTRO–01–0145–FEDER–000007). Authors have benefited from the participation in the ISSI workshops.

**Disclosure of Potential Conflicts of Interest** The authors declare that they have no conflicts of interest.

**References**

Balmaceda, L.A., Solanki, S.K., Krivova, N.A., Foster, S.: 2009, A homogeneous database of sunspot areas covering more than 130 years, J. Geophys. Res. 114, A07104. DOI: 10.1029/2009JA014299.

Baranyi, T.; Győri, L.; Ludmány, A.: 2016, On-line Tools for Solar Data Compiled at the Debrecen Observatory and Their Extensions with the Greenwich Sunspot Data, Solar Phys. 291, 3081. DOI: 10.1007/s11207-016-0930-1.

Bonifácio, V.: 2017, Costa Lobo (1864-1945), the Coimbra Spectroheliograph and the Internationalisation of Portuguese Astronomy, History of Astronomy in Portugal, Imprimerie Centrale de l'Université de Nantes, Nantes, pp. 113-138.

Carrasco, V.M.S., Vaquero, J.M., Aparicio, A.J.P., Gallego, M.C.: 2014, Sunspot Catalogue of the Valencia Observatory (1920 – 1928), Solar Phys. 289, 4351. DOI: 10.1007/s11207-014-0578-7.

Carrasco, V.M.S., Lefèvre, L., Vaquero, J.M., Gallego, M.C.: 2015, Equivalence relations between the Cortie and Zürich sunspot group morphological classifications, Solar Phys. 290, 1445. DOI: 10.1007/s11207-015-0679-y.

Chatzistergos, T., Usoskin, I.G., Kovaltsov, G.A., Krivova, N.A., Solanki, S.K.: 2017, New reconstruction of the sunspot group numbers since 1739 using direct calibration and "backbone" methods, Astron. Astrophys. 602, A69. DOI: 10.1051/0004-6361/201630045.

Clette, F., Lefèvre, L.: 2016, The New Sunspot Number: Assembling All Corrections, Solar Phys. 291, 2629. DOI: 10.1007/s11207-016-1014-y.

Clette, F., Svalgaard, L., Vaquero, J.M., Cliver, E.W.: 2014, Revisiting the Sunspot Number. A 400-year perspective on the solar cycle, Space Sci. Rev. 186, 35. DOI: 10.1007/s11214-014-0074-2.

Cliver, E.W., Ling, A.G.: 2016, The Discontinuity Circa 1885 in the Group Sunspot Number, Solar Phys. 291, 2763. DOI: 10.1007/s11207-015-0841-6.

da Costa Lobo, F.M.: 1929, Introdução, Anais do Observatório Astronómico da Universidade de Coimbra, Universidade de Coimbra, Coimbra.

Erwin, E.H., Coffey, H.E., Denig, W.F., Willis, D.M., Henwood, R., Wild, M.N.: 2013, The Greenwich Photo-heliographic Results (1874 – 1976): Initial Corrections to the Printed Publications, Solar Phys. 288, 157. DOI: 10.1007/s11207-013-0310-z.

Hoyt, D.V., Eddy, J.A.: 1982, An atlas of variations in the solar constant caused by sunspot blocking and facular emissions from 1874 to 1981, National Center for Atmospheric Research, NCAR Technical Note TN−194+STR, Boulder, Colorado.

Hoyt, D.V., Schatten, K.H.: 1998, Group sunspot numbers: A new solar activity reconstruction. Solar Phys. 179, 189. DOI: 10.1023/A:1005007527816.

Lefèvre, L., Clette, F.: 2014, Survey and Merging of Sunspot Catalogs, Solar Phys. 289, 545. DOI: 10.1007/s11207-012-0184-5.

Lefèvre, L., Aparicio, A.J.P., Gallego, M.C., Vaquero, J.M.: 2016, An Early Sunspot Catalog by Miguel Aguilar for the Period 1914 – 1920, Solar Phys. 291, 2609. DOI: 10.1007/s11207-016-0905-2.

Leonardo, A.J.F., Martins, D.R., Fiolhais, C.: 2011, Costa Lobo and the Study of the Sun in Coimbra in the First Half of the Twentieth Century, J. Astron. Hist. Herit. 14, 41.

Lockwood, M., Owens, M.J., Barnard, L.A., Usoskin, I.G.: 2016, An Assessment of Sunspot Number Data Composites over 1845-2014, Astrophys. J. 824, 54. DOI: 10.3847/0004-637X/824/1/54.

Mandal, S., Hegde, M., Samanta, T., Hazra, G., Banerjee, D., Ravindra, B.: 2017, Kodaikanal digitized white-light data archive (1921-2011): Analysis of various solar cycle features, Astron. Astrophys. 601, A106. DOI: 10.1051/0004-6361/201628651.

Maunder, E.W.: 1904, Note on the distribution of sun-spots in heliographic latitude, 1874 to 1902. Mon. Not. Roy. Astron. Soc. 64, 747.

Mouradian, Z., Garcia, A.: 2007, Eightieth Anniversary of Solar Physics at Coimbra, The Physics of Chromospheric Plasmas ASP Conference Series 368, Astronomical Society of the Pacific, San Francisco.

Owens, B.: 2013, Long-term research: Slow science, Nature 495, 300. DOI: 10.1038/495300a.

Svalgaard, L., Schatten, K.H.: 2016, Reconstruction of the Sunspot Group Number: The Backbone Method, Solar Phys. 291, 1. DOI: 10.1007/s11207-015-0815-8.

Usoskin, I.G.: 2017, A history of solar activity over millennia, Living Rev. Solar Phys. 14, 3. DOI: 10.1007/s41116-017-0006-9.

Usoskin, I.G., Kovaltsov, G.A., Lockwood, M., Mursula, K., Owens, M., Solanki, S.K.: 2016, A New Calibrated Sunspot Group Series Since 1749: Statistics of Active Day Fractions, Solar Phys. 291, 1. DOI: 10.1007/s11207-015-0838-1.

Vaquero, J.M., Svalgaard, L., Carrasco, V.M.S., Clette, F., Lefèvre, L., Gallego, M.C., Arlt, R., Aparicio, A.J.P., Richard, J.-G., Howe, R.: 2016, A Revised Collection of Sunspot Group Numbers, Solar Phys. 291, 3061. DOI: 10.1007/s11207-016-0982-2.

Willis, D.M., Coffey, H.E., Henwood, R., Erwin, E.H., Hoyt, D.V., Wild, M.N., Denig, W.F.: 2013a, The Greenwich Photo-heliographic Results (1874 – 1976): Summary of the Observations, Applications, Datasets, Definitions and Errors, Solar Phys. 288, 117. DOI: 10.1007/s11207-013-0311-y.

Willis, D.M., Henwood, M.N., Wild, M.N., Coffey, H.E., Denig, W.F., Erwin, E.H., Hoyt, D.V.: 2013b, The Greenwich Photo-heliographic Results (1874 – 1976): Procedures for Checking and Correcting the Sunspot Digital Datasets, Solar Phys. 288, 141. DOI: 10.1007/s11207-013-0312-x.

Zolotova, N.V., Ponyavin, D.I., Marwan, N., Kurths, J.: 2009, Long-term asymmetry in the wings of the butterfly diagram, Astron. Astrophys. 503, 197. DOI: 10.1051/0004-6361/200811430.

Zolotova, N.V., Ponyavin, D.I., Arlt, R., Tuominen, I.: 2010, Secular variation of hemispheric phase differences in the solar cycle, Astron. Nachr. 331, 765. DOI: 10.1002/asna.201011410.